\documentclass[graybox,natbib]{svmult1}
\usepackage{amssymb}
\usepackage{mathptmx}       
\usepackage{helvet}         
\usepackage{courier}        
\usepackage{type1cm}        
\usepackage{makeidx}         
\usepackage{graphicx}        
\usepackage{multicol}        
\usepackage[bottom]{footmisc}
\makeindex             
\begin{document}
\title*{ Collapse and fragmentation of Gaussian barotropic protostellar clouds}
\author{F. G\'omez-Ram\'irez$^{\dagger\ddagger}$,
        J. Klapp$^{\ddagger +}$,
Jorge L. Cervantes-Cota$^{\ddagger *}$, 
        G. Arreaga-Garc\'ia$^{**}$, 
        D. Bahena$^{\S}$ 
         }
\authorrunning{F. G\'omez-Ram\'irez, J. Klapp,   Jorge L. Cervantes-Cota,  G. Arreaga-Garc\'ia, D. Bahena } 
\institute{ . \at $^\dagger$ Facultad de Ciencias, Universidad Aut\'onoma del Estado de M\'exico,
El Cerrillo Piedras Blancas, C.P. 5200, Estado de M\'exico, M\'exico. 
\at
$^\ddagger$  Instituto Nacional de Investigaciones Nucleares, Carretera M\'exico-Toluca S/N La Marquesa, 
Ocoyoacac,  C.P. 52750, Edo. de M\'exico.  
\at
$^*$  Berkeley Center for Cosmological Physics, LBNL , Berkeley, California 94720, USA.
\at
$^{**}$
Centro de Investigaci\'on en F\'isica de la Universidad de Sonora, A. P. 14740,  Hermosillo, C.P. 83000, Sonora, M\'exico.
\at
$^\S$ 
Astronomical Institute of the Academic sciences, Bo\u{c}m\'{\i} II 1401, 14130 Praha 4, Czech Republic.
\at
$^+$  jaime.klapp@inin.gob.mx
}


%
%
\maketitle

\abstract{\newline 
We examine the problem of the collapse and fragmentation of molecular
clouds with a Gaussian density distribution with high resolution, double precision numerical simulations using
the GADGET-2 code.  To describe the thermodynamic pro\-per\-ties of the cloud during  the collapse
-to mimic the rise of temperature predicted by radiative transfer-
we use a barotropic equation of state that introduces a critical density to separate the isothermal
and adiabatic regimes.  We discuss the effects of this cri\-ti\-cal density in the formation of multiple
systems.  We confirm the tendency found for Plummer and Gaussian models that if the collapse changes from
isothermal to adiabatic at earlier times that occurs for the models with a lower critical density, the collapse
is slowed down, and this enhances the fragments' change to survive. However, this effect happens up to a
threshold density  below which single systems tend to form.  On the other hand, by setting a
bigger initial perturbation amplitude, the collapse is faster and in some cases a final single object is formed.
}

\section{Introduction}
\label{INTRO:1}
The protostellar objects that begin their main sequence and pre-main sequence are mainly distributed in
binary and multiple systems that  suggests they were formed during the process of collapse and fragmentation
of molecular clouds with dense cores and gas envelopes, see  \citet*{sk01a,tohline} and references
therein.   In recent years several authors have  considered
different realizations of molecular clouds to study  their collapse and fragmentation.    Although most
fragmentation calculations apply to initially uniform conditions, see for instance \citet*{bodeniv}, it is clear from 
the observations that molecular cloud cores are centrally condensed \citet*{ward,andre98,motte98}.   Thus,  a number 
of collapse models starting from centrally condensed, Gaussian density 
profiles have also been made. A particular computationally demanding 
isothermal, Gaussian cloud model was first calculated in \citet*{boss91}, and thereafter recalculated by other authors as 
a further test case to check both the likelihood of fragmentation during the isothermal collapse phase and the reliability 
of the numerical code results \citet*{bb96,truelove,boss98,boss2000,sk01b,sk01c}.

So far the great majority of this research has concentrated upon the early phases of star formation,
when the collapse is dynamical, first isothermal and then non-isothermal. However, precise knowledge of the
dependence of temperature on density at the transition from isothermal to nonisothermal
collapse requires solving the radiative transfer problem coupled to a fully self-consistent energy equation.  However,
the full non-isothermal computation represents a severe computational burden imposed by solving the radiative transfer
equations at high spatial resolution, even in the Eddington approximation. Therefore, it has been common  to
use instead a simple barotropic equation of state \citet*{boss2000} that clearly simplifies the computational problem and  it
turns out to be a good approximation for the dynamical collapse of the molecular clouds, see \citet*{Arreaga2007,Arreaga2008}.
In the present work, a barotropic equation of state is assumed to simulate
the transition from isothermal to adiabatic collapse. The motivation of this study is to investigate the sensitivity of
fragmentation to the effects of thermal retardation by varying the value of the critical density at which nonisothermal
heating is assumed to begin.  In \citet*{Arreaga2007,Arreaga2008}  it is  studied the evolution of a Gaussian
density profile, and found that
by diminishing the critical density it enhances the fragmentation. A similar result was found for Plummer models in \citet*{Arreaga2010}.

The present work is a continuation of the analysis in  \citet*{Arreaga2007,Arreaga2008}   in which we employ a 
Gaussian density profile and perform the same type of numerical computations but now using double precision 
in the GADGET-2 code.  Particularly, we study the effect of varying the critical density of the barotropic equation 
of  state in the collapse and fragmentation of the molecular protostar.  We also 
analyze the effect to increasing the initial perturbation amplitude.

\section{Initial conditions and collapse models}
\label{IniticalConditions:1.1}

According to astronomical observations, the regions from which
stars are formed consist basically of molecular hydrogen clouds
at a temperature of $\sim 10\, K$. Therefore, the ideal equation of state
is a good approximation to account for the thermodynamics of
the gas in these clouds.  The cloud models are based  on the 
standard isothermal test case, as in the
variant considered in \citet*{bb93}.  However, once gravity has produced a
substantial contraction of the cloud, the opacity increases, the
collapse changes from isothermal to adiabatic and the gas begins
to heat. To include this rise in temperature into our
calculations, we use the barotropic equation of state proposed in  \citet*{boss2000}.

    In order to correctly describe the non-isothermal regime, one
needs to solve the radiative transfer problem coupled to the
hydrodynamic equations, including a fully self-consistent energy
equation to obtain a precise knowledge of the dependence of
temperature on density. The implementation of radiative transfer
has already been included in some mesh-based codes. In SPH,
the incorporation of radiative transfer has in general not been
very satisfactory, perhaps with the exception of reference \citet*{whitehouse}, in
which they used the f-limited diffusion approximation to model 
the collapse of molecular cloud cores.
These authors suggested that there are important differences in
the temperature evolution of the cloud when radiative transfer is
properly taken into account.

 However, after comparing the results of  the simulations  performed by  \\
\cite{Arreaga2007,Arreaga2008} with those of  reference
\cite{whitehouse} for the uniform density cloud, it is concluded that
the barotropic equation of state in general behaves quite well
and that we can capture the essential dynamical behavior of the
collapse. The simulations in this work are consequently carried
out using the following barotropic equation of state:

\begin{equation}
 p=c_{iso}^2\rho+K\rho^\gamma,
\end{equation}
where $\gamma$ is the adiabatic exponent in the opacity thick regime and $K$ is a constant set by
$ K=c_{iso}^2\rho_{crit}^{1-\gamma}$, where $\rho_{crit}$ defines the critical density above which the collapse
changes from isothermal to adiabatic, and for a molecular hydrogen
gas the ratio of specific heats is $\gamma = 5/3$, because we only
consider translational degrees of freedom.

With the above prescriptions, the local sound speed becomes
\begin{equation}
 c=c_{iso}\left[ 1+\left( \frac{\rho}{\rho_{crit}}\right) ^{\gamma-1}\right] ^{1/2},
\end{equation}
so that $c\approx c_{iso}$ when $\rho\ll \rho_{crit}$ and  $c\approx C_{ad}=\gamma^{1/2} c_{iso}$.

The molecular cloud collapse simulations in this work begin with initial conditions in accordance to the thermodynamic model proposed in
\cite{bb93}.  Accordingly, the models start with a spherical cloud of mass  $M=1 M_{\odot}$, radius
$R=4.99 \times 10^{16} \, {\rm cm} \sim 0.016 \, {\rm pc}$, and at a tempe\-ra\-ture $T=10\, {\rm K}$.
The initial model is composed by an ideal gas with an average molecular weight $\mu \sim 3$.   We have chosen the initial sound speed
$c_{iso}$  and  angular velocity   $\omega_0$ in such a way that for all models the initial ratio of thermal and rotational energies to gravitational energy
are such that $\alpha=E_{\rm therm}/|E_{\rm grav}| \sim   0.26$ and $\beta=E_{\rm rot}/|E_{\rm grav}| \sim 0.16$.  The gas isothermal sound speed
is  $c_{iso} \sim  1.90 \times 10^{4}\, {\rm cm}\, {\rm s}^{-1}$, and the average free fall time  is  $5.10\times 10^{11}$s.  Additionally, we impose a
small perturbation to the density profile of the following form:
\begin{equation}  \label{DENSIDAD:1}
 \rho=\rho_0\left[ 1+ a \, cos(m\phi)\right] ,
\end{equation}
where $m$ is an integer number, $\phi$  the azimuthal angle around the $z-$axes, and $a$ is the perturbation amplitude.

The chosen density profile is Gaussian with the above-mentioned initial conditions, as in \citet*{Arreaga2007,sk01a}:
\begin{equation}
 \rho\left( r\right) = \rho_c \exp{\left[
                                          -\left( \frac{r}{b}\right)^2
                                   \right],}
\end{equation}
where $\rho_c=1.7\times10{-17}\, {\rm gr}\,{\rm cm}^{-3}$  is the initial central density and $b\approx 0.578 R$ is
a length chosen such that the density is 20 times smaller there.    On the other hand,
solid-body rotation is assumed  at the rate of $\omega_0=1.0\times 10^{-12}s^{-1}$.

\section{Numerical methods}
\label{NUMERICALMETHOD:1}
The computations of this work were performed using the parallel  code GADGET-2, which is described in full in
\cite{gadget2}. The code is suitable for studying isolated, self-gravitating systems with high spatial
resolution.  The code is based on the tree-PM methods for computing the gravitational forces and on standard SPH methods
for solving the 3D Euler hydrodynamics equations.  For a review on the theory and applications of SPH we refer the reader
to \citet*{monagan2005}.

In order to set up the initial particle distribution, we first define a Cartesian box with
sides equal to twice a specified radius $R_b \gtrsim R=4.99\times 10^{16}$ cm, and with  its geometrical center coinciding with
the origin ($x=y=z=0$) of a Cartesian coordinate system. The box is then subdivided into regular cubics
cells of volume $\Delta^3=\Delta x \Delta y \Delta z$ each. The spherical cloud is then copied within
the box by placing an SPH particle in the center of each cell at distances $d \le R$ from the origin,
so that the region outside the sphere is empty. A little amount of disorder is added to the regular
distribution of particles by shifting each particles a distance $\Delta/4$ from its
cell-center location and along  a specified direction, which is chosen randomly among the three Cartesian axes.
We defined the mass of particle $i$ at location ($x_i,\,y_i,\,z_i$) to be $m_i=\rho(x_i,\,y_i,\,z_i)\Delta^ 3,$
where
\begin{equation}
\rho(x_i,\,y_i,\,z_i)= \rho_c \exp{\left[- \frac{x_i^2 +y_i^2+z_i^2}{b^2}\right]} .
\end{equation}\label{DENSIDAD:2}

Solid-body rotation about the $z-axis$ is assumed in a counter clockwise sense by assigning
to particle $i$ an initial velocity given by $v_i=(\omega_0 x_i,-\omega_0 y_i,0).$
Finaly, the bar mode density perturbation given by equation (\ref{DENSIDAD:1}) applied by
modifying the mass of particle $i$ according to $m_i\rightarrow m_i[1+a \cos{(m\phi_i)}],$
where $\phi_i$ denotes the azimuthal position of that particle.  The computations were performed in the parallel
cluster of the National Institute of Nuclear Research-Mexico, equipped with 28 AMD Quad-core
(64 bits) Opteron Barcelona processors.

\section{Results} 

In this section we present the results obtained using the GADGET-2 code of the co\-llap\-se and 
fragmentation of Gaussian molecular clouds.  The collapse of  the Gaussian cloud first calculated 
in \citet*{boss91}, later by other authors \citet*{bb96,truelove,boss98,sk01b,sk01c,Arreaga2007}  using high 
spatial resolution, and in the present work we perform the same computations 
as in \citet*{Arreaga2007}, but adding double precision and using $10^{7}$  SPH particles in each simulation.

\begin{figure}
\includegraphics[angle=0,width=3.8cm]{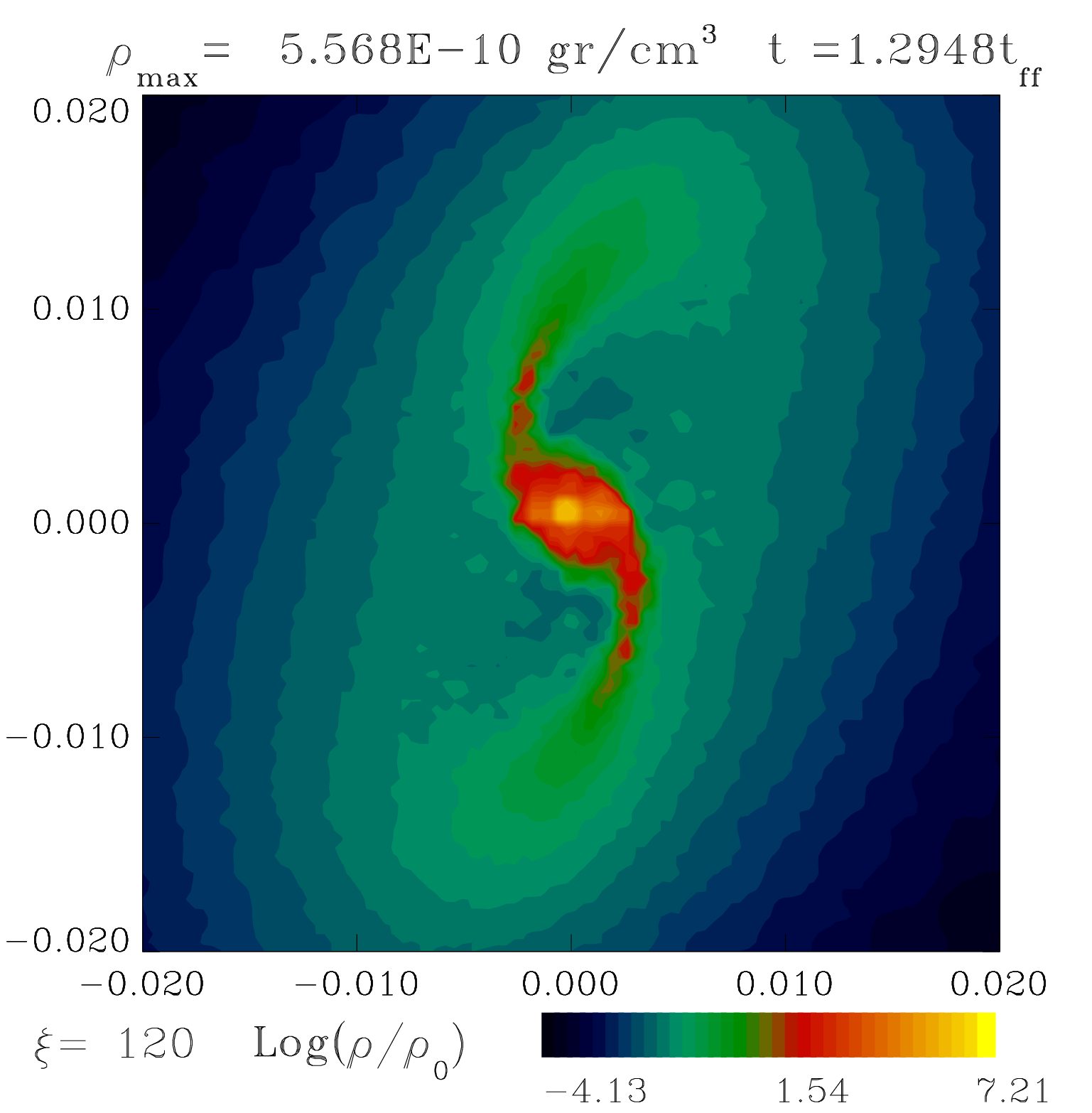}
\includegraphics[angle=0,width=3.8cm]{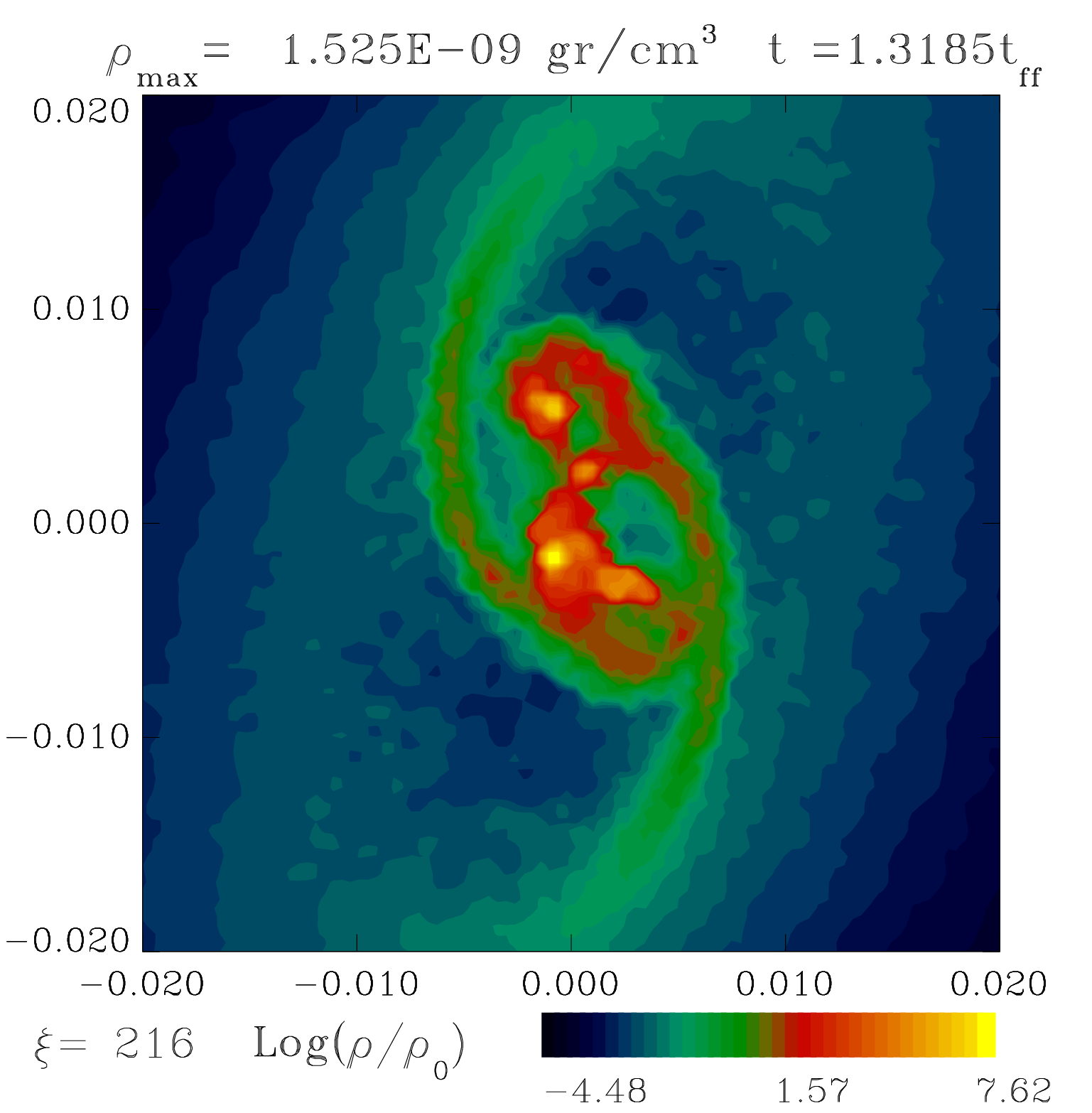}
\includegraphics[angle=0,width=3.8cm]{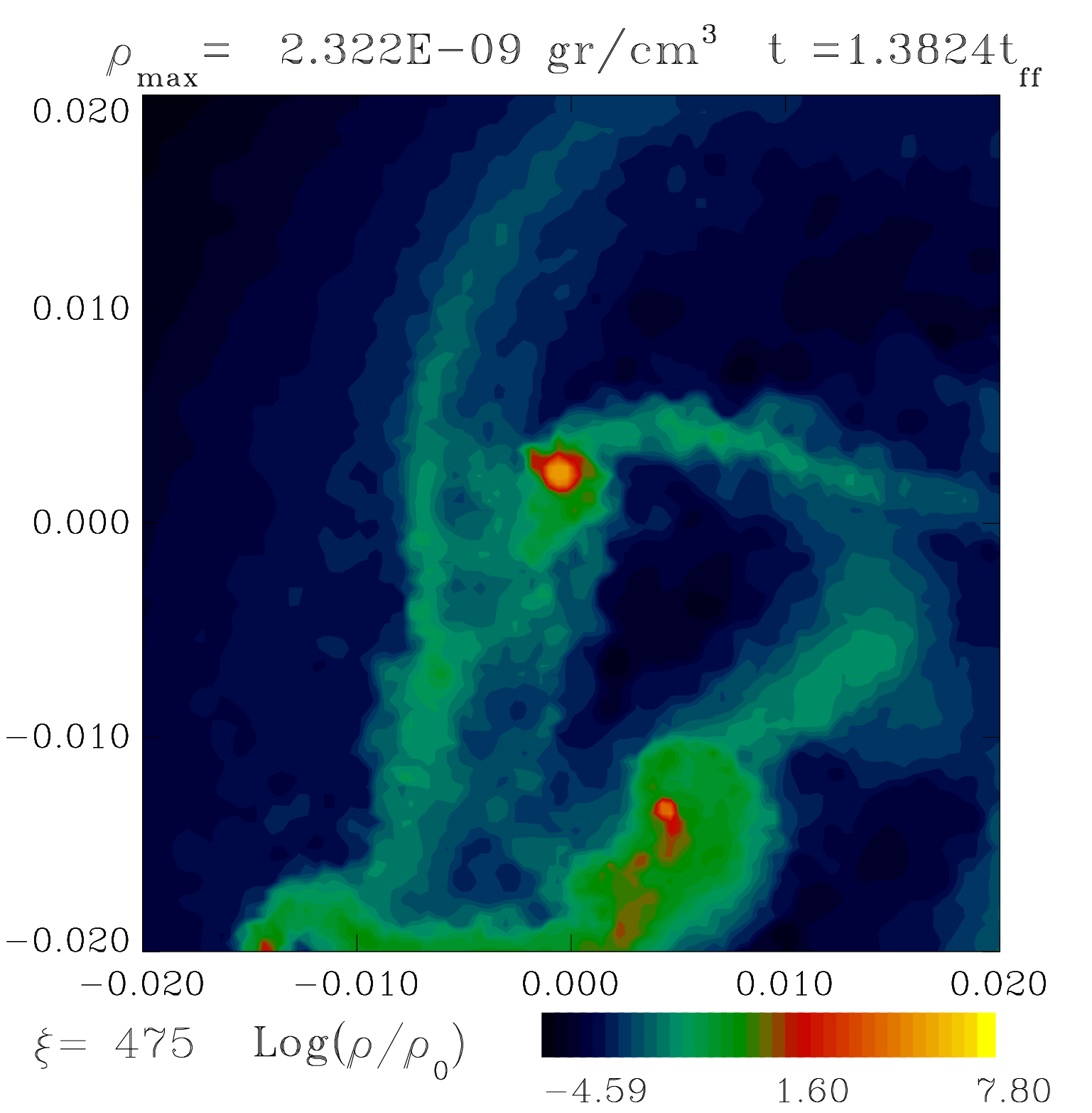}
\caption{Iso-density contour plots  at the equatorial plane of model G6A1 for three different times.}  \label{ModelG6A1}
\end{figure}

\begin{figure}
\includegraphics[angle=0,width=3.8cm]{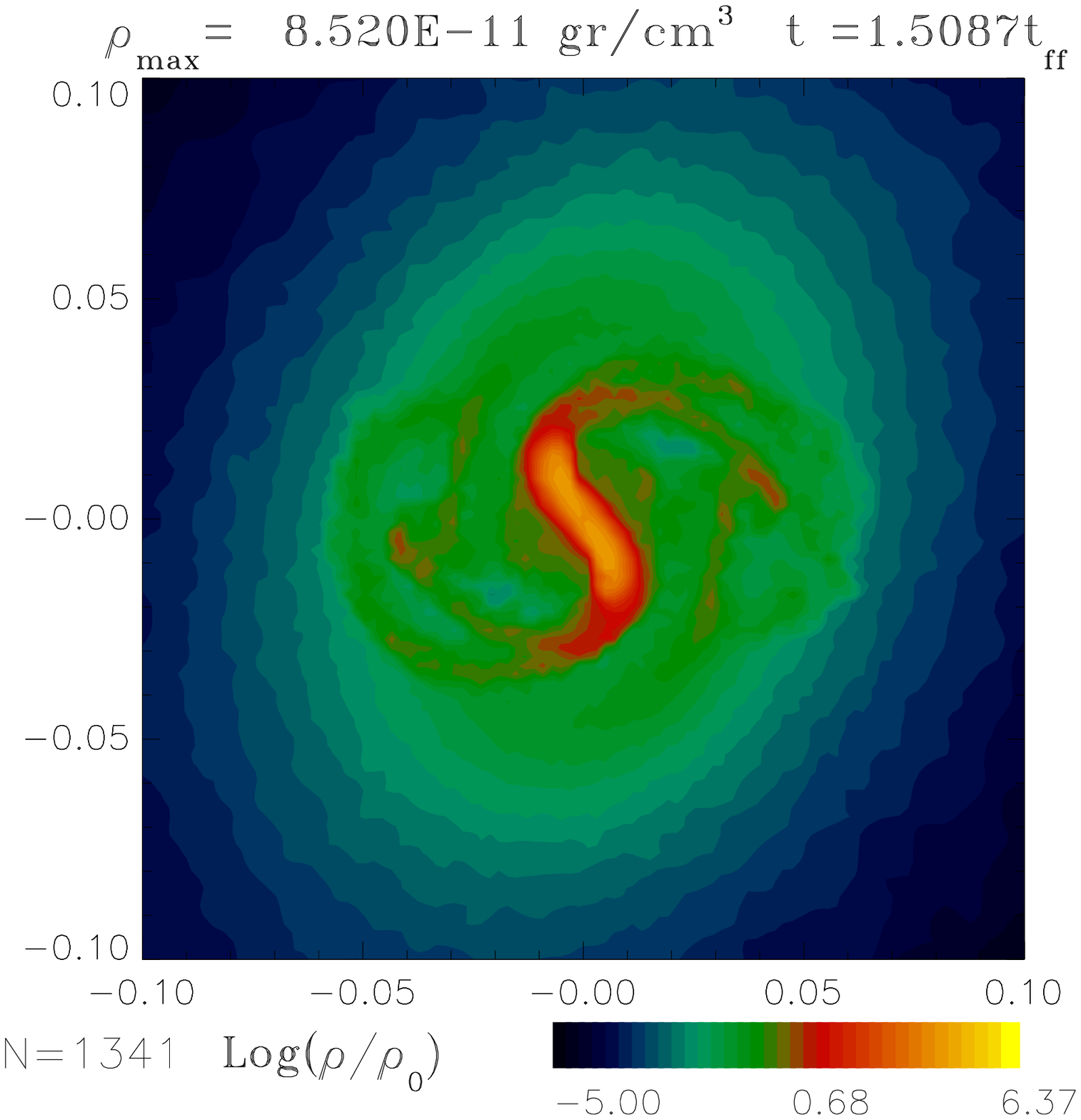}
\includegraphics[angle=0,width=3.8cm]{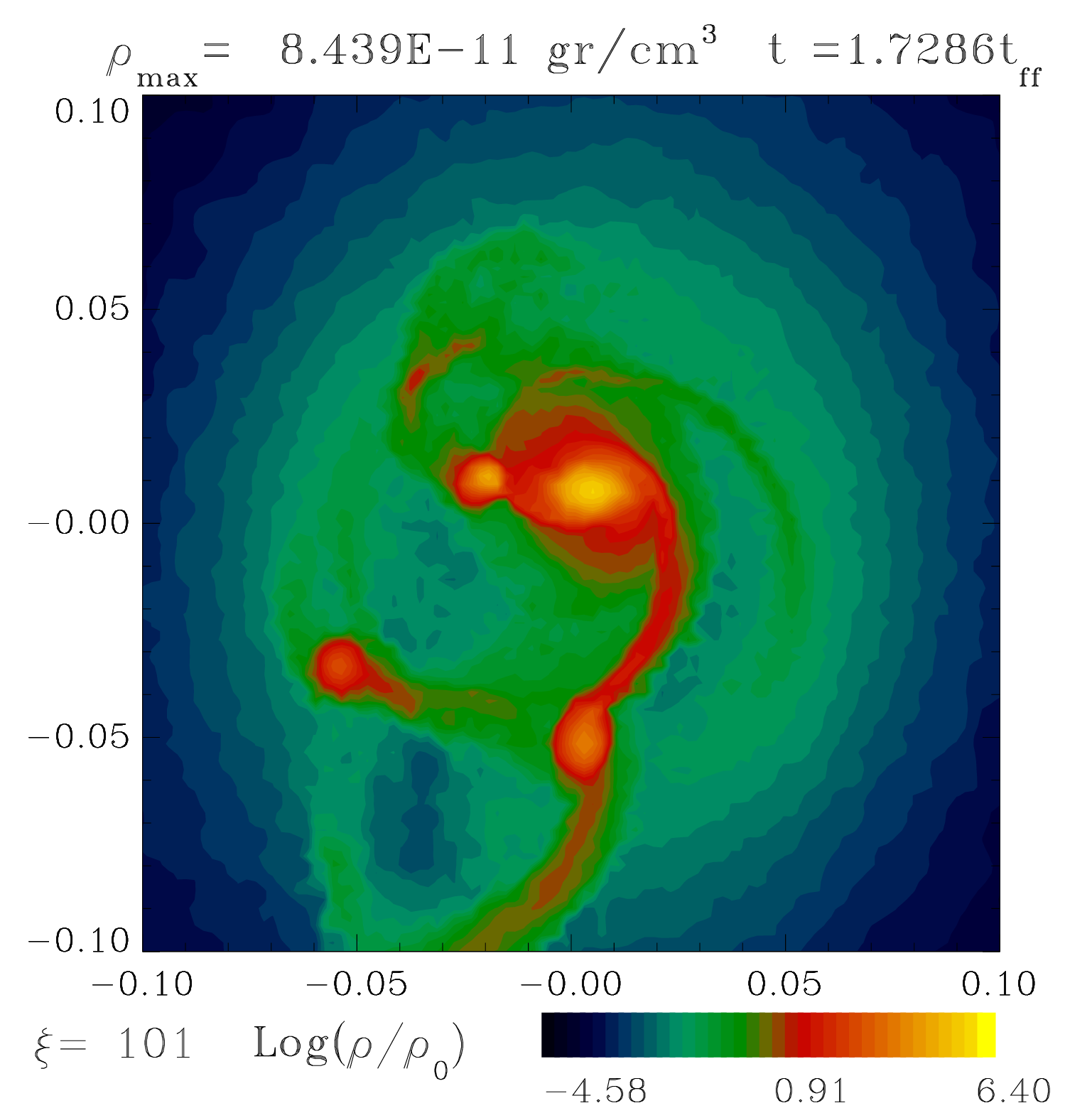}
\includegraphics[angle=0,width=3.8cm]{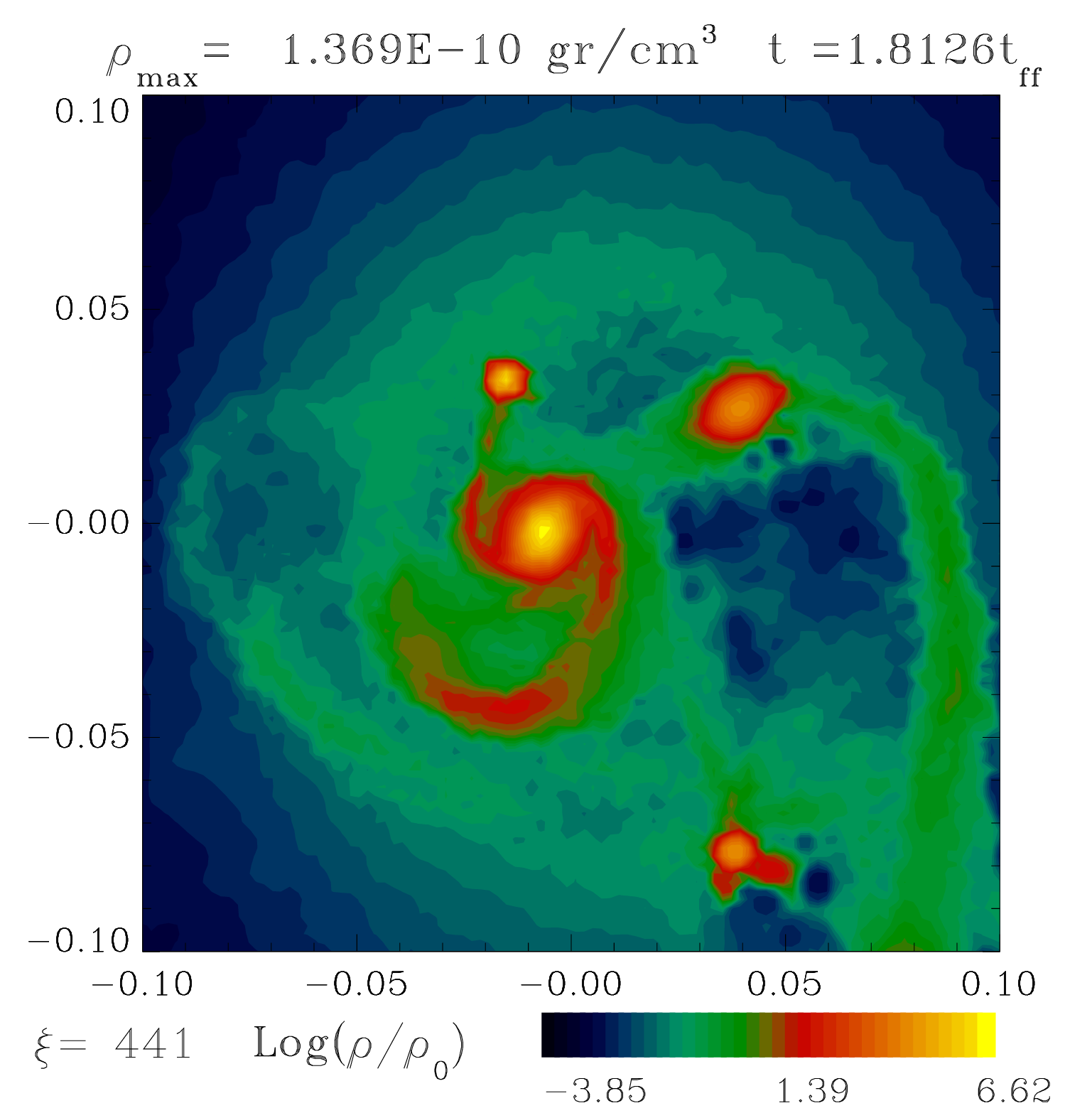}
\caption{Iso-density contour plots  at the equatorial plane of model G6B1 for three different times.}  \label{ModelG6B1}
\end{figure}

\begin{figure}
\includegraphics[angle=0,width=3.8cm]{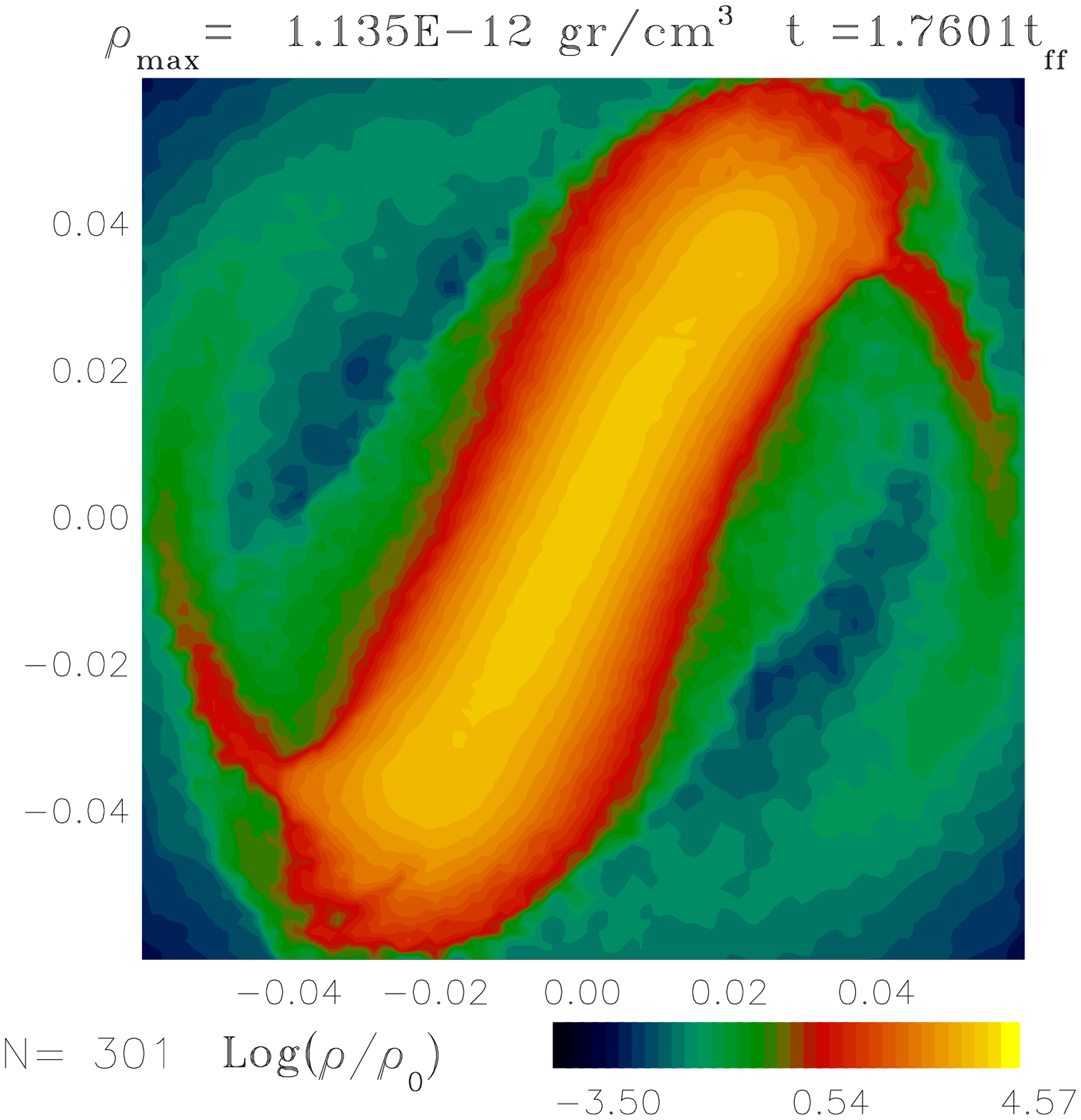}
\includegraphics[angle=0,width=3.8cm]{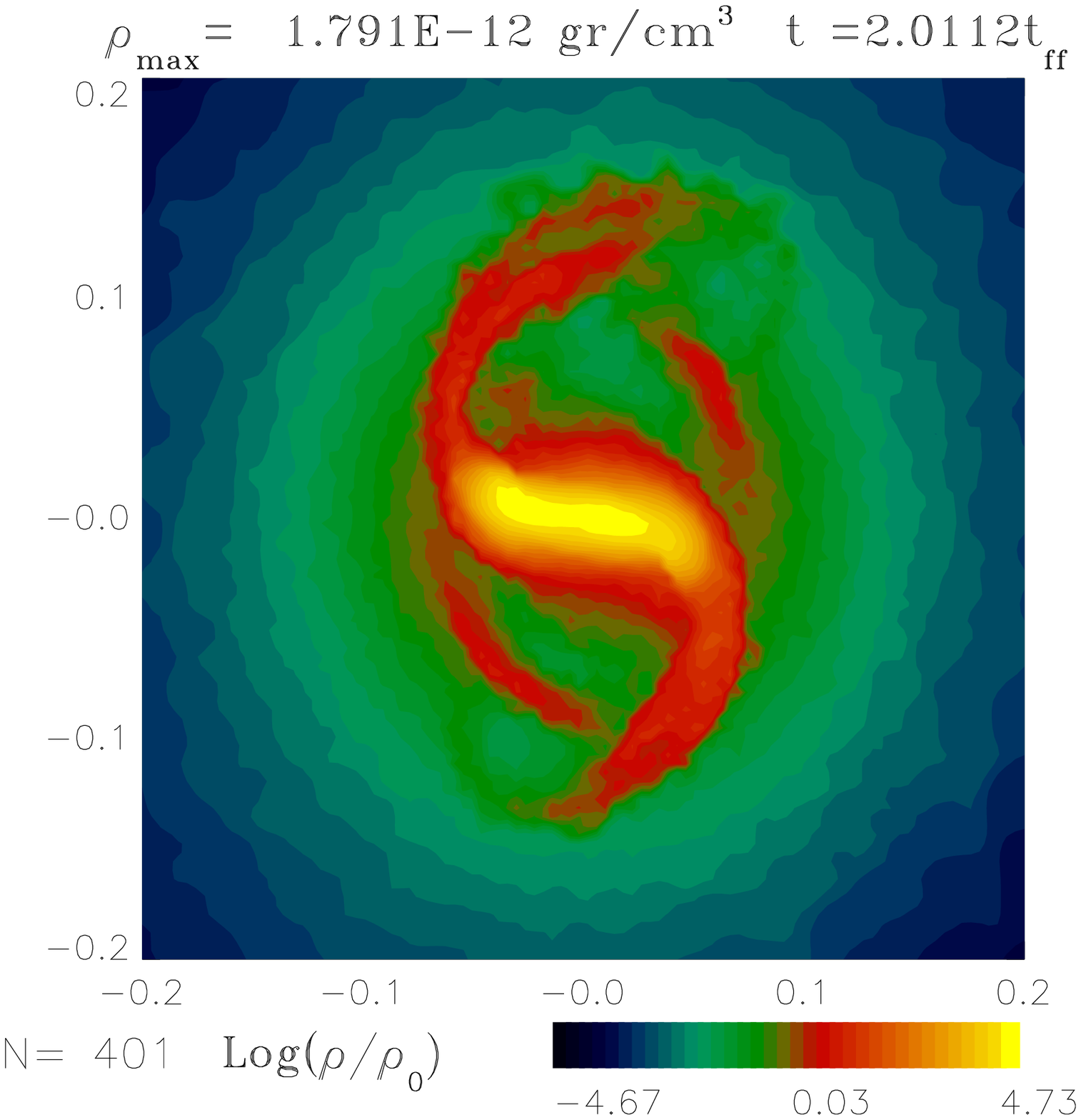}
\includegraphics[angle=0,width=3.8cm]{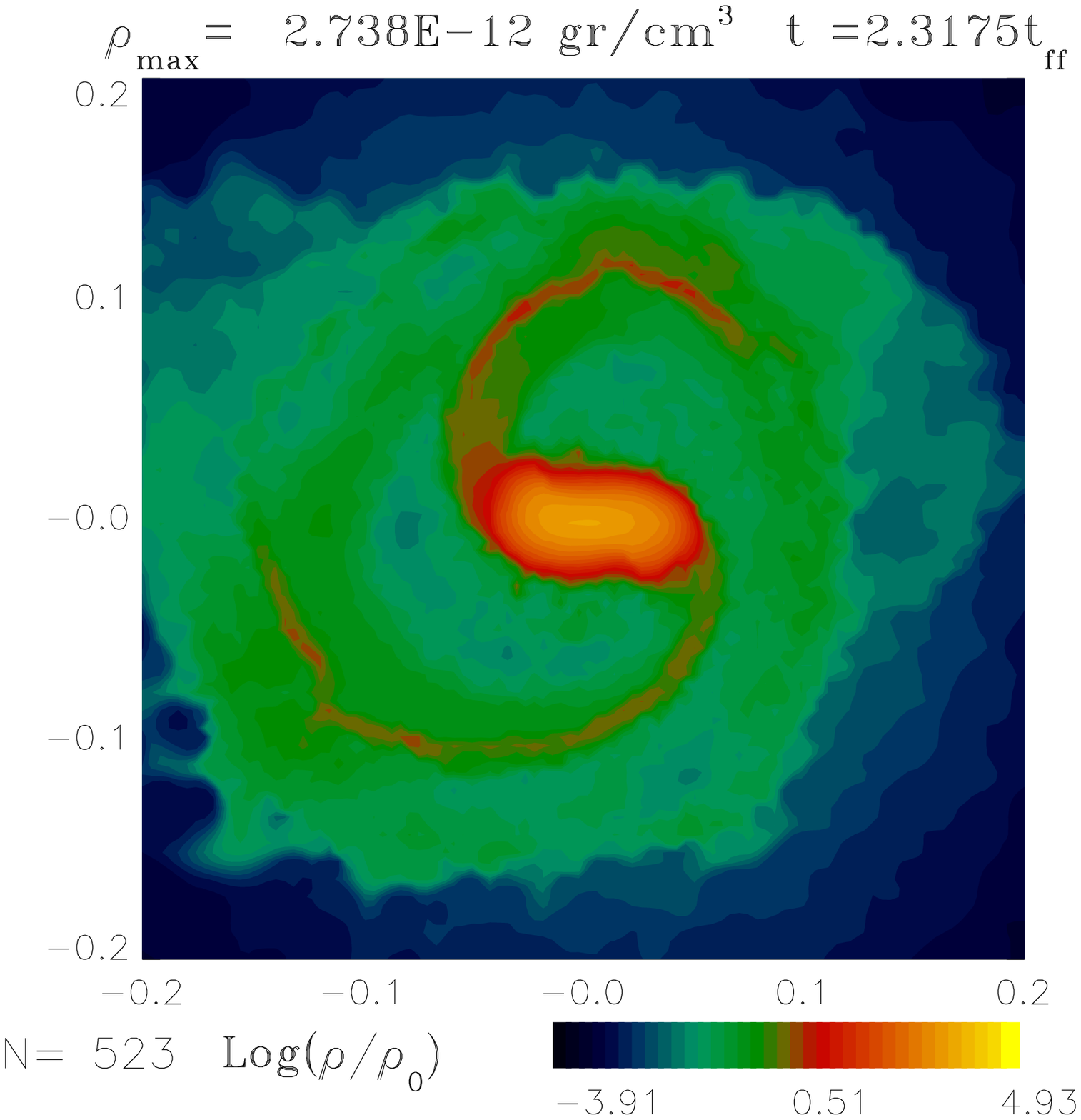}
\caption{Iso-density contour plots  at the equatorial plane of model G6C1 for three different times.}  \label{ModelG6C1}
\end{figure}

\begin{figure}
\includegraphics[angle=0,width=3.8cm]{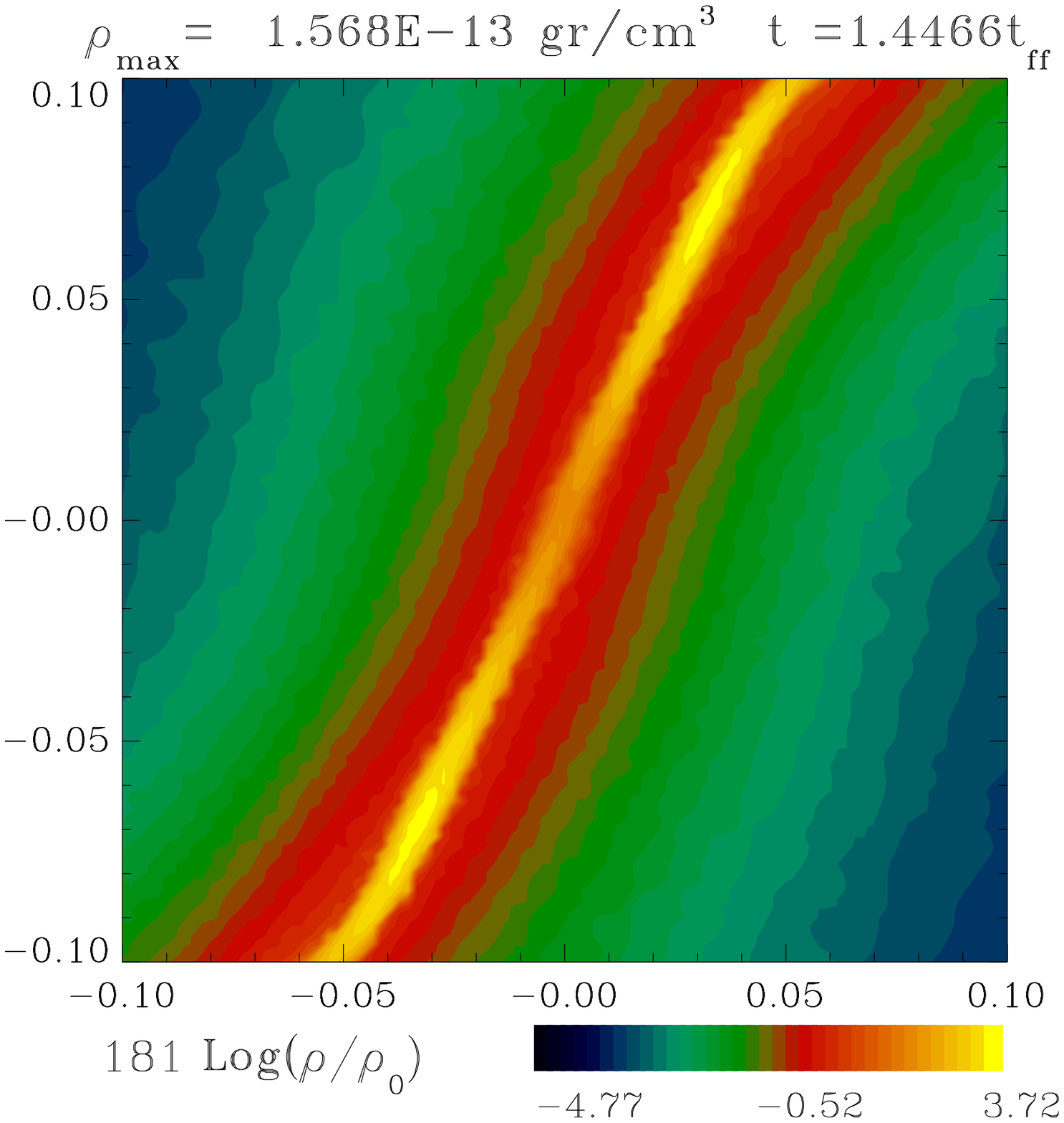}
\includegraphics[angle=0,width=3.8cm]{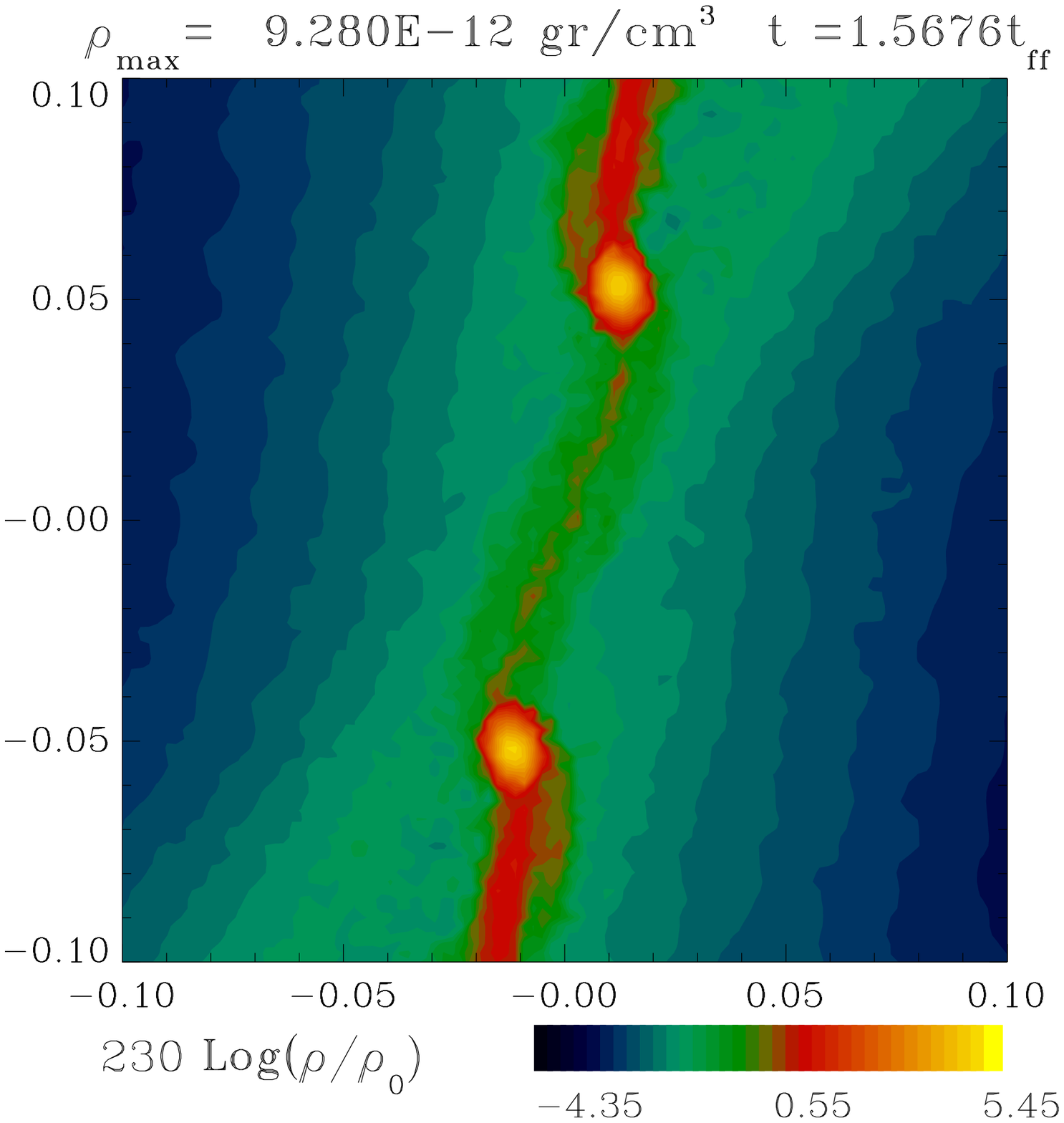}
\includegraphics[angle=0,width=3.8cm]{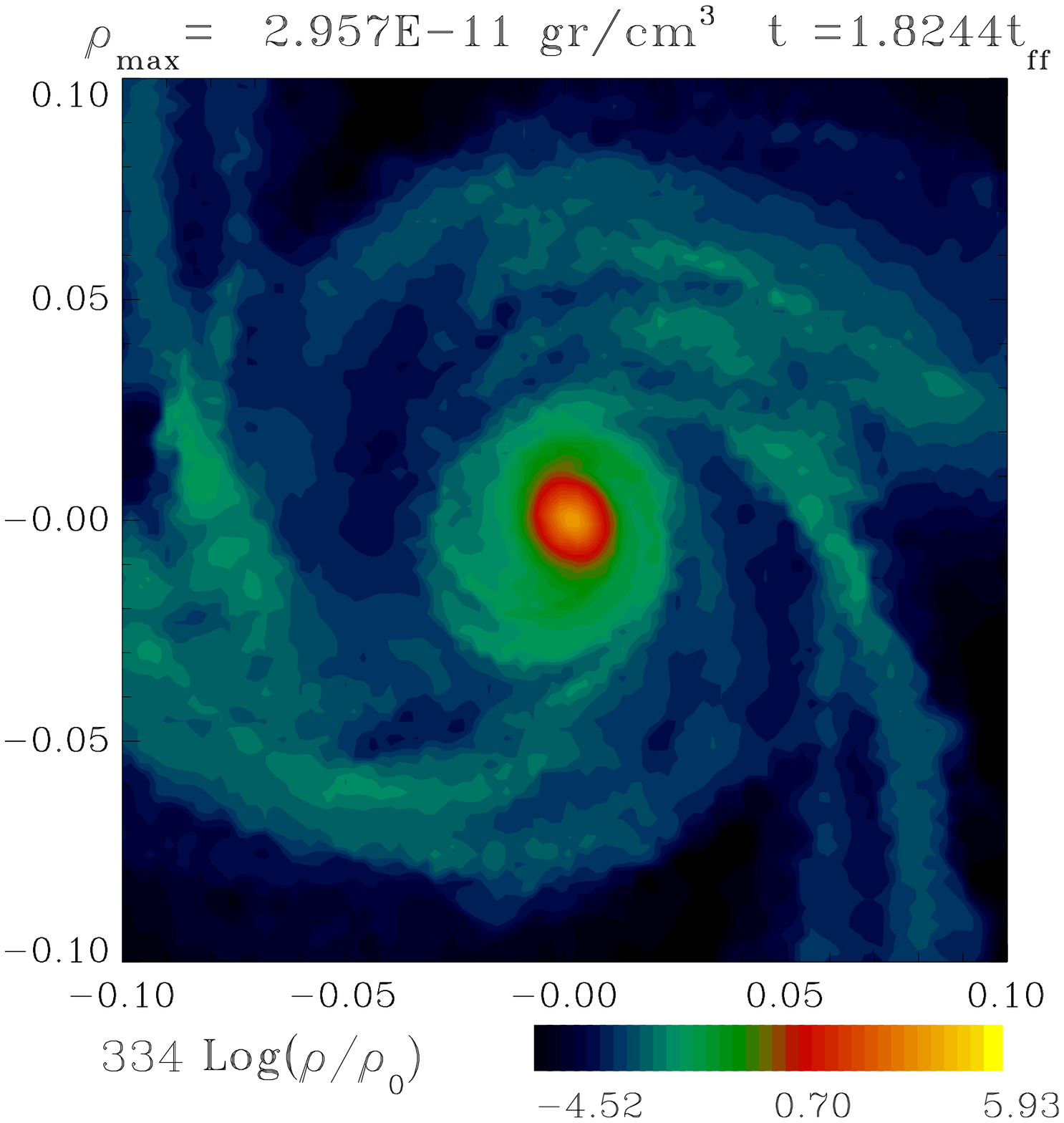}
\caption{Iso-density contour plots  at the equatorial plane of model G6B5 for three different times.}   \label{ModelG6B5}
\end{figure}

We present four different cases for the barotropic collapse with the parameters as shown in table \ref{tabla:1}. The results are
illustrated with iso-density contour plots for a slide at the equatorial plane of the cloud in 
figures  \ref{ModelG6A1}-\ref{ModelG6B5}.  The bar located at the bottom of the plots shows the ${\rm log}_{10}$ density 
range at a time  $t$ and normalized with the initial central 
density  $\rho_c$. A color scale is then associated with the value of ${\rm log}_{10}$.  For instance, the color scale
uses yellow to indicate higher densities, blue for lower densities, and green and orange for intermediate densities.

The free fall time $t_{ff}\approx \sqrt{3 \pi/(32 G \rho_c})$ sets a characteristic time scale for the collapse of protostellar clouds which is given in terms of the
central density  and it is the same for all models, see for instance \citet*{Arreaga2010}.

\begin{table} 
\caption{Gaussian collapse models. The model types are explained as follows: The letter G refers to Gaussian, the number 6 refers to ten
millions particles, the letters
A, B, and C distinguish among the different critical densities, and finally the last digit refers to the amplitude of the initial perturbation.}
\label{tabla:1}       
%
%
\begin{tabular}{p{1.cm}p{2cm}p{2.7cm}p{2cm}p{4.9cm}}
\hline\noalign{\smallskip}
Model & $\rho_{crit}\,({\rm g}\,{\rm cm}^{-3})$ &  Amplitude $a$  & Final outcome  \\
\noalign{\smallskip}\svhline\noalign{\smallskip}
G6A1 & $5\times 10^{-12}$ &  0.1 & Binary\\
G6B1 & $5\times 10^{-14}$ &   0.1 & Quadruple\\
G6C1 & $5\times 10^{-15}$ &   0.1 & Single\\
G6B5 & $5\times 10^{-14}$ &   0.5 & Single\\
\noalign{\smallskip}\hline\noalign{\smallskip}
\end{tabular}
\end{table}

The aim of the present models is to explore, using double precision variables, the sensibility of fragmentation to both the effect of thermal
retardation  due to nonisothermal  heating that is controlled by the critical density,  and the effect to increase of perturbation amplitude.

The construction of the initial conditions ensures that the collapse initial stage is similar for all models considered here.
The initial phase of the collapse proceeds in such a way that the material falls down to the rotation plane,
while material near the central midplane undergoes a weak expansion perpendicular to the rotation axis, causing the formation of two
overdense blobs from the initial $m=2$ perturbation seed.  At about the end of the first free-fall time, the expansion stops
and the middle region begins to collapse, the blobs fall toward the center and merge to form a prolate structure. By this time, the overall cloud
has already been compressed into a flat disk with an inner bar that begins to rotate.

We begin the discussion with models G6A1, G6B1, seen in figures  \ref{ModelG6A1} and \ref{ModelG6B1}.   The model G6A1
collapses faster than the corresponding single precision model reported
in  \citet*{Arreaga2007}, but the overall dynamics looks similar.  We now proceed to decrease the critical density, as shown in model G6B1.  We find
however important differences  in comparison to  \citet*{Arreaga2007}, which reports  as an 
end product a binary system formed by a transient quadrupole system, but we obtain as a final product a stable quadrupole system.
The fact that by diminishing the critical density  the fragmentation enhances, found for Gaussian \citet*{Arreaga2007,Arreaga2008}
and  Plummer models \citet*{Arreaga2010}, seems to be reproduced for models  G6A1 and G6B1.  However,  by diminishing even more the
critical density in model G6C1, see table \ref{tabla:1}, the collapse and fragmentation is slowed down and
the fragmentation is less favored, as  seen in figures  \ref{ModelG6B1} and  \ref{ModelG6C1}.

The model G6B5, shown in figure \ref{ModelG6B5},  has a bigger initial perturbation amplitude  
that provokes that systems collapse
earlier to form a filamentary structure, which evolves to a transient binary system that later collapses in a single system. The final
system ejects significant amount of gas out of the core. In comparison the same model  but with a smaller
amplitude, model G6B1,  forms a stable quadrupole system.

\section{Conclusions}

In this work, we have followed the early phases of cloud collapse and fragmentation up to the formation of the
proto stellar core using the GADGET-2 code with high spatial resolution and double precision, using $10^7$ SPH
particles. The initial conditions for the cloud models are chosen to be the standard isothermal test case, as in the
variant considered in \citet*{bb93}, but for a centrally condensed, Gaussian cloud, that was first treated in \citet*{boss91}, and
further considered by other authors \citet*{Arreaga2007}. The main  results are summarized as follows:

By augmenting from single to double precision, the collapse happens earlier and the number of end products augments,
as seen in our simulations  in comparison the same models in \citet*{Arreaga2007}.

On the other hand, we find that the effect of diminishing the critical density of the barotropic equation of state, provokes
the collapse to slow down, and this enhances the fragments' change to survive. However, this effect happens up to a  threshold density, as seen in
our simulation G6C1, where considered a low critical density, $\rho_c = 5 \times 10^{-15} {\rm g}/{\rm cm}^3$, and  a single system was formed.

Moreover, models  with a bigger initial perturbation amplitude provoke that systems collapse  
earlier, and in some cases, in  the form of a  filamentary structure which evolves to a transient binary system that later collapses in a single system.

\section*{Acknowledgements }  
J.L.C.C. thanks the Berkeley Center for Cosmological Physics for hospitality, and gratefully acknowledges
support from a UC MEXUS-CONACYT Grant, and a CONACYT Grant No. 84133-F. 


\end{document}